\documentclass[fleqn,usenatbib]{mnras}

\usepackage[T1]{fontenc}
\usepackage{ae,aecompl}
\usepackage{newtxtext,newtxmath}

\usepackage{multicol}        
\usepackage{supertabular,booktabs}	
\usepackage{lipsum}	
\usepackage{fmtcount}
\usepackage[normalem]{ulem} 
\usepackage{graphicx, amsmath, amssymb, epsfig, natbib}
\usepackage{bm}
\usepackage{fmtcount}
\usepackage{commath}	
\usepackage{epstopdf}
\usepackage{ctable}	
\usepackage{longtable}	
\usepackage{multirow}		
\usepackage{pdflscape}
\usepackage{footnote}	

\def\msol{\hbox{$\hbox{M}_\odot$ }}

\newcommand{\kms}{km s$^{-1}$}
\newcommand{\wat}{H$_2$O }
\newcommand{\dee}{$^\circ$}
\newcommand{\meth}{CH$_{3}$OH }
\newcommand{\am}{\ensuremath{^\prime} }
\newcommand{\as}{\ensuremath{^{\prime\prime}} }
\newcommand{\+}{$\pm$}
\newcommand{\mum}{$\mu$m}
\newcommand{\RNum}[1]{\uppercase\expandafter{\romannumeral #1\relax}}	

\title[6.7 GHz \meth maser survey of the CMZ]{A 6.7 GHz methanol maser survey of the central molecular zone}

\author[Mr. Rickert et al.]{Matthew Rickert$^{1,2}$, F. Yusef-Zadeh$^{1}$, and J. Ott$^{2}$\\
$^{1}$CIERA	and	Department of Physics and Astronomy, 
Northwestern University, 	2145 Sheridan Rd., 	Evanston, IL 60208-3112, USA\\
$^{2}$National Radio Astronomy Observatory, 1003 Lopezville Rd,	Socorro, NM 87801}

\date{Accepted XXX. Received \today; in original form 9 August 2018}

\pubyear{2018}

\begin{document}		
\label{firstpage}
\pagerange{\pageref{firstpage}--\pageref{lastpage}}
\maketitle
\begin{abstract}

The Central Molecular Zone (CMZ) spans the inner $\sim$450 pc ($\sim$3\dee) of our Galaxy.  This 
region is defined by its enhanced molecular emission and contains 5\% of the entire Galaxy's molecular gas 
mass.  However, the number of detected star forming sites towards the CMZ may be low for the amount 
of molecular gas that is present, and improved surveys of star formation indicators can help clarify 
this. With the Karl G Jansky Very Large Array (VLA), we conducted a blind survey of 6.7 GHz methanol (\meth) 
masers spanning the inner 3\dee $\times$ 40\am (450 pc $\times$ 100 pc) of the Galaxy.  
We 
detected 43 \meth masers towards 28 locations,   $\sim$16 of which 
are new 
detections. The velocities of most of these masers are consistent with being located within the CMZ.  
A majority of the detected \meth masers are distributed towards positive Galactic longitudes, similar 
to 2/3 of the molecular gas mass distributed  at positive Galactic longitudes.
The 6.7 GHz 
\meth maser is  an excellent  indicator  of high mass ($>8\msol$) star formation, with new detections indicating  sites of massive star formation in the CMZ.

\end{abstract}

\begin{keywords}
	Galaxy: center---Galaxy: nucleus---star formation --- catalogs --- surveys
\end{keywords}

\section{Introduction} \label{sec:intro}

The Central Molecular Zone (CMZ) occupies  the inner $\sim$ 450 pc ($\sim$ 3\dee) of the Galaxy.  This 
region is defined by enhanced molecular gas emission. While being less than 1\% of the Galaxy's 
volume, it is estimated to contain 5\% of the Galaxy's molecular gas mass (\citealt{xu15}; 
\citealt{dahmen98}).  The ISM in the CMZ  is unique in its physical properties, 
containing  a higher 
molecular gas density (\citealt{ao13}), higher temperature (\citealt{ao13}; \citealt{oka05}; 
\citealt{martin04}; \citealt{morris96}), and increased turbulent velocity (\citealt{ao13}; \citealt{lis98}; 
\citealt{martin97}; \citealt{morris96}; \citealt{bally87}) than elsewhere in the Galaxy.

How star formation progresses in the unique environment of the CMZ is a current research topic.  While the CMZ contains a relatively high star formation rate (SFR) of $\sim0.1$\msol 
yr$^{-1}$, the large molecular gas mass calls into question the efficiency of star formation in this environment 
(\citealt{barnes17}; \citealt{longmore13}). \cite{longmore13}  show that the fraction of the 
number of star formation indicators per fraction of dense gas is lower in the CMZ than in the Galactic disk, 
implying  that  the star formation is suppressed in the CMZ.  \cite{far09} instead have shown that the CMZ 
falls along the Kennicut-Schmidt relation, indicating that the CMZ star formation is neither 
suppressed nor enhanced.  While one of the ways to address this issue is to better constrain the 
molecular gas mass in the CMZ, we have instead chosen to address this issue by obtaining improved 
counts of the number of star forming sites in the CMZ.

Due to 30 magnitudes of  visual extinction in the direction of the CMZ, masers are useful probes of 
star formation in the CMZ. We conducted the first blind interferometric survey of 6.7 GHz methanol (\meth) masers 
towards the CMZ using the Jansky Very Large Array (VLA). This maser is a strong indicator of high 
mass ($>8\msol$) star formation (\citealt{minier03}; \citealt{dish98}).  Unlike other \meth masers 
(i.e 36, 44, 96 GHz), the 6.7 GHz \meth maser is radiatively pumped by photons from dust heated by 
high mass stars (\citealt{menten91s}). The 6.7 GHz \meth maser is solely associated with star 
formation (\citealt{breen13}), unlike other masers that are collisionally pumped (i.e. 
36, 44, 96 GHz \meth and 22 GHz \wat) that can also be produced by  shocks from clump-clump collisions. 

Many of the previous surveys of young stellar objects (YSOs) cited in the works of 
\cite{longmore13} and \cite{far09} have utilized single dish telescopes and/or targeted surveys (i.e. 
\citealt{walsh11,chambers14} and \citealt{caswell11}), which could cause the survey to miss detections.  We 
instead use a blind interferometric survey with a large and uniform spatial coverage  
to obtain a sensitivity limited sample of masers.

\section{Observations}\label{sec:obs:meth}
We conducted a single survey of 6.7 GHz \meth masers with the Karl G. Jansky Very Large Array (VLA\footnote{The VLA is 
operated by the National Radio Astronomy Observatory (NRAO), which is a facility of the National Science Foundation 
(NSF) operated under cooperative agreement by Associated Universities, Inc.}) on September 
9, 2015. 
  The VLA was in its A-array configuration providing a spatial resolution of 0.83\as $\times$ 
0.29\as. Table \ref{tab:obs:meth} summarizes the observing setup, including the spectral resolution and 
velocity coverage.

\begin{figure*}\label{fig:methpointings}
	\includegraphics[width=\textwidth]{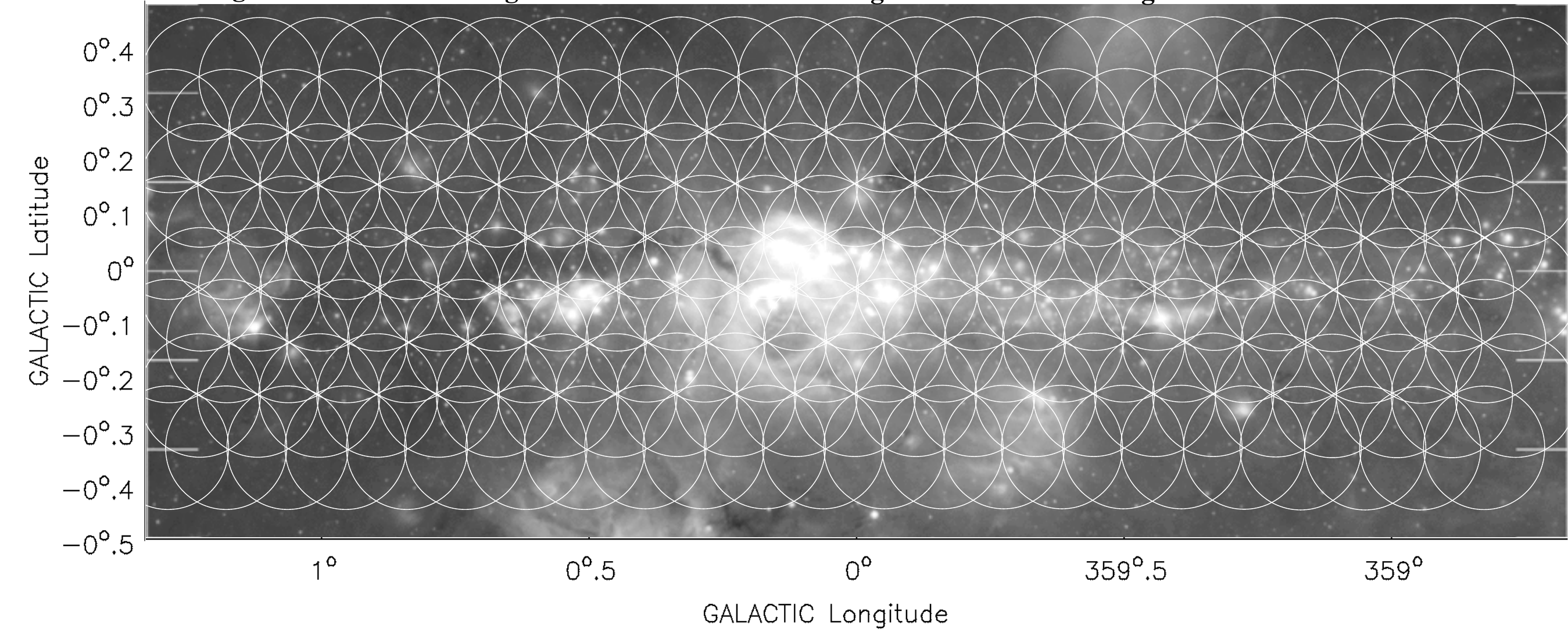}
	\caption{The observed CH$_3$OH fields are plotted on a 24 $\mu$m~map. Each circle is roughly 6' wide, which corresponds to the primary beam at 6 GHz. At the top are labeled some prominent sources at their corresponding longitudes.}
\end{figure*}

The survey covers the inner $3^{\circ}\times40\am$ ($l \times b$) of the GC using  203 overlapping 
pointings  with a standard Nyquist sampling to ensure near uniform sensitivity.  
Figure \ref{fig:methpointings} shows these pointings over plotted on a 24 \mum map (\citealt{far09}).
Each pointing
 was observed for 47 seconds and a total of $\sim10$ minutes was spent 
on calibrators. J1733-1304 was 
observed as a bandpass and gain calibrator and 3C286 was observed as a flux and bandpass calibrator. The initial 
desired noise level (based on the online VLA exposure calculator) of 0.01 Jy beam$^{-1}$ was almost obtained for a 
majority of the resulting map, with the average RMS being 0.08 Jy beam$^{-1}$ (Table \ref{tab:gaus:meth} later shows 
the channel RMS near each of the detections).

\subsection{Comparison to Previous Blind 6.7 GHz  \meth Maser Surveys of the CMZ}\label{sec:comp:meth}
There have  been two other blind surveys of 6.7 GHz \meth masers that have covered the CMZ.  Table \ref{tab:methsurv:comp} compares 
 our survey (indicated as Rickert2018) to these two other blind surveys.
All other 6.7 GHz \meth maser surveys  have 
either targeted specific sources or did not cover the CMZ.  This is in part due to 
 the lack of 6.7 GHz receivers on VLA and GBT  that could observe this frequency in the past\footnote{The current VLA C-band receivers (4--8 GHz) were not available on all the antennas before 2013 (\citealt{cband})}. 
 
The most recent 
untargeted survey of 6.7 GHz \meth masers covering the CMZ is the methanol multibeam (MMB) survey (\citealt{green09}).  The MMB survey used the Parkes single dish telescope for their initial detections (although later followed 
up these detections with 
the Australia Telescope Compact Array (ATCA)).  Our \meth survey has a comparable sensitivity and higher initial resolution than 
the MMB survey (\citealt{caswell96a}).

Before the MMB survey, \cite{chambers09} also conducted a blind 6.7 GHz \meth maser ATCA survey of the CMZ. 
Due to the configuration they were in, they had a poorer resolution (2\as $\times$ 4\as, 0.18 \kms) than the follow up 
ATCA observations used in the MMB survey.  
Although the spectral resolution of \cite{chambers09} is better than our VLA observations, 
their spatial resolution and sensitivity are both poorer than ours.  

\subsection{Data Reduction and Imaging}\label{sec:obs:im:meth}

To reduce  the large data set, 
the primary visibility file was split into mosaics of 7 fields in a Nyquist sampled 
mosaic pattern. This ensures that the central field contains the full survey sensitivity.  Common Astronomy Software 
Applications (CASA, \citealt{CASA}) was used as the primary reduction, calibration, imaging, and analysis package for 
this survey. The split visibility files were then continuum subtracted (using the CASA task \textit{uvcontsub}) with 
line free channels near $\sim\pm$250 \kms\ selected for fitting the continuum. Each field was imaged individually (with a cleaning threshold of 0.055 Jy and a maximum 200 iteration limit with a loop gain of 0.1) over just the central $\sim$ 230 \kms\ (as 
opposed to the full \+ 390 \kms\ ) and were then combined together using CASA's image.linearmosaic function into hexagonal patterns of 7 fields each. The resulting images were each 6750 $\times$ 6750 pixels (with a 0.1\as pixel size 
and a 0.86\as $\times$ 0.29\as synthesized beam) with 570 channels (each 0.4 \kms, 8 kHz wide).

\subsection{Maser Identification}\label{sec:id}

Masers candidates were initially identified by creating peak maps of the central field of the 7 field mosaic. The CASA function \textit{ia.findsource} was then used to 
identify all sources above 5 times the theoretical RMS (this resulted in initially a large number of false detections that were removed in a later step).  Each 
 maser candidate had a Gaussian fitted to its spectrum. This was done to  prevent single channel detections and to distinguish between spectral peaks.  
If there were any sources near\footnote{Within 0.015\dee\ and the FWHM of the Gaussian fitted to each source's spectrum} each other in position and velocity then only the single maser candidate with the highest peak flux density was kept as a detection.  The 
RMS was then measured by averaging the RMS from $\sim$ 5 regions (each 4\as wide and located within 1\am of the maser) 
within the channel of the spectral peak (this was to account for how the RMS may not be uniform throughout the 
survey).  Finally, any sources whose fitted peak flux densities fell below 5 times this measured RMS were then removed, 
leaving a sensitivity limited catalog of sources that we were confident were masers. A similar method to this was used by \cite{cotton16}, 
except that they used integrated maps, masked the channels before collapsing the image cube into an integrated map, 
and they did not fit Gaussians to their spectra.

\section{Results}\label{meth:res}

We detected 43 \meth maser candidates towards 28 distinct locations\footnote{Compact sources $\sim$ 1\as wide.}. 
 Table \ref{tab:gaus:meth} gives the parameters of the  Gaussian  fit 
 to each maser's spectrum.  
Columns 1-10 of Table \ref{tab:gaus:meth} respectively give: the source ID\footnote{The ID is in order of decreasing Galactic longitude.}, the 
Galactic coordinates, right ascension (RA), declination (Dec), peak flux density, RMS\footnote{The RMS 
	was measured in the channel of the peak maser emission from $\sim$ 5 regions (each 4\as wide and located within 1\am 
	of the maser) and then averaged together.}, central velocity, 
FWHM, 
integrated line intensity, and  associated sources\footnote{The associated sources are  prominent  features of the CMZ (i.e. Sgr 
	A, Sgr B, etc.), references to sources from other catalogs (the footnotes indicate source type, catalog references, and the maximum distance between the \meth maser and the catalog source), and indicators of new detections. }.  
The Gaussian fit uncertainties are given for the flux density, velocity, FWHM, and integrated flux density.

Figure \ref{methspec} shows the spectra for all 
the 43 \meth masers. Each spectrum is labeled with the source ID and Galactic coordinates from Table 
\ref{tab:gaus:meth}.  Within each spectrum, the identified spectral peaks are labeled with their corresponding 
velocities (from Table \ref{tab:gaus:meth}).  Some of the spectra (such as those for sources 12, 20, 25, and 28) have 
apparent spectral peaks that do not have identified velocities.  This is generally because: there is a brighter maser 
at the same velocity within 0.015\dee (and thus the unlabeled peak was deemed a potential artifact), the peak was 
below our detection limit, or a Gaussian could not be fitted to the spectrum (because it is only 1-2 channels wide). 
The channel width for these spectra\footnote{Due to not having a narrower channel width, some of the bright and narrow 
masers display Gibbs ringing.
Most modern interferometers record the visibilities as a function of time lag (and not frequency), and then Fourier 
transform to obtain the visibilities as a function of frequency, which can thus result in the Gibbs phenomenon 
appearing in the spectra (\citealt{bigbook})
} is 0.4 \kms. 

\subsection{Brightness Temperature}\label{sec:bright}

In order to verify the non-thermal nature of the detected \meth maser candidates and support that they are masers, we determined the
brightness temperature to distinguish them from thermal \meth emission.  The brightness temperature ($T_b$) is determined by:
\begin{equation}
T_{b} = \frac{c^{2} S}{2 k \nu^{2} \Omega},
\label{eq:brightnesstemp}
\end{equation}
where $c$ is the speed of light, $k$ is the Boltzmann constant, $\nu$ is the frequency (6.66852 GHz), and $\Omega$ is 
the solid angle of the source. As these sources are spatially unresolved, we use a solid angle ($\Omega$) equal to the 
synthesized beam (0.86\as $\times$ 0.29\as).  $S$ is the flux density in the peak channel 
integrated over the 
synthesized beam.  This calculated brightness temperature is thus a lower limit on the actual brightness temperature, 
as it ignores the other channels that make up the spectrum and overestimates the angular size of the source.  The second to last column of Table 
\ref{tab:gaus:meth} lists these calculated minimum brightness temperatures (and the propagated error). The $T_b$ values span $\sim2\times10^{3}$ to $\sim3\times10^{6}$ K.  The molecular gas in the CMZ has temperatures $<$700 K.  Given the high brightness temperatures and narrow line widths ($<$2.9 \kms, Table \ref{tab:gaus:meth}), 
the detected maser candidates are consistent with being masers. 

\section{Discussion}

\begin{figure*}

	\includegraphics[width=\textwidth]{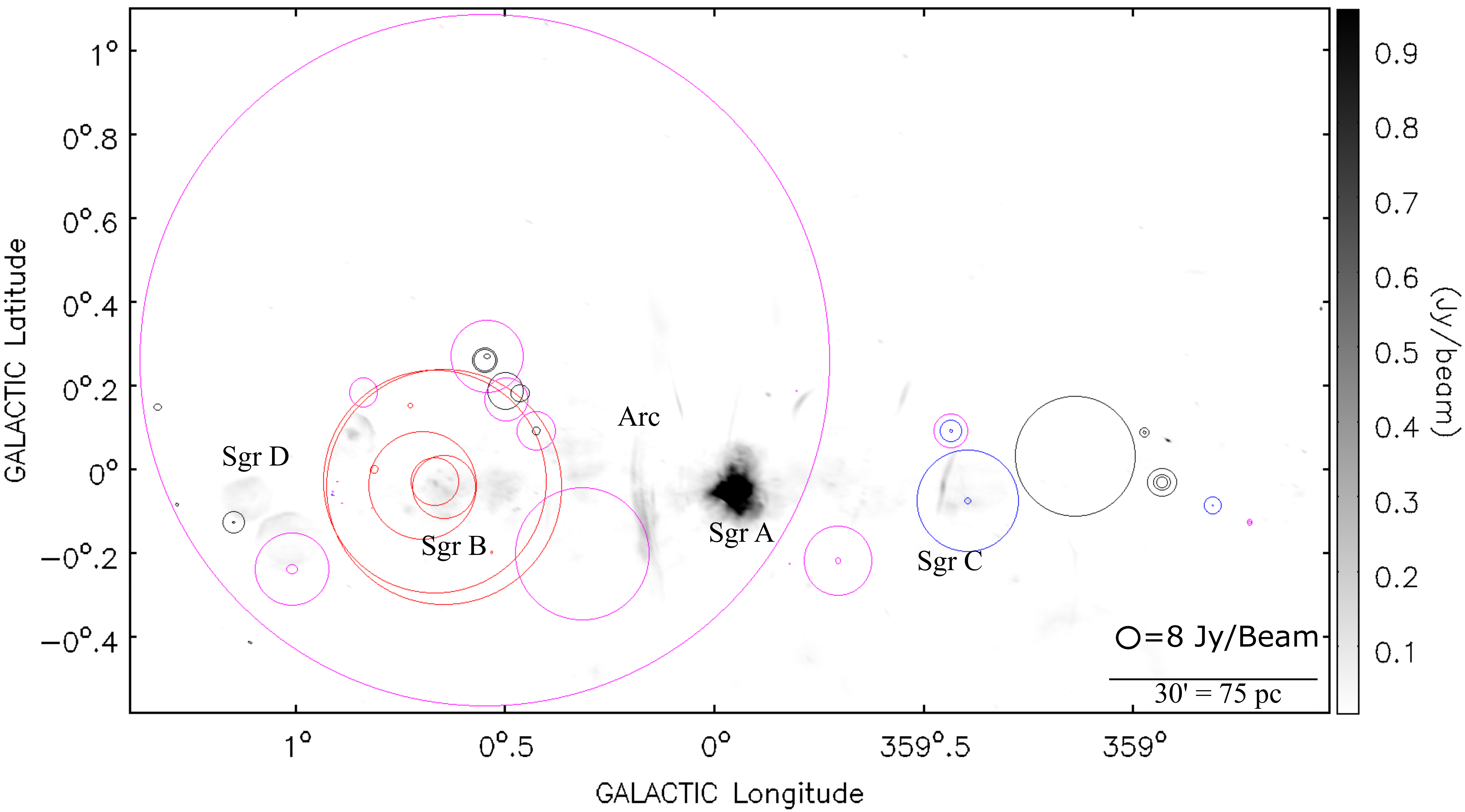}
	\caption{The detected \meth masers have been plotted on a 90cm continuum map (\protect\citealt{law08}). The size of the circles is proportional to the maser peak flux density. The color of the circles corresponds to the maser velocity: cyan $\ge$100 \kms$>$ red $\ge$30 \kms\ $>$ magenta $\ge$ 0 \kms\ $>$ black  $\ge$-30 \kms\ $>$ blue $\ge$ -100 \kms\ $>$ green.}\label{methfancy}	
\end{figure*}

The positive Galactic longitudes of the CMZ contain two to three times more molecular gas mass 
than  at negative longitudes (\citealt{launhardt02}). Similar to the molecular gas, more 6.7 GHz \meth masers are located towards positive Galactic longitudes, with 24 of the \meth masers appearing towards positive Galactic longitudes and 
only 10 towards negative longitudes.  In order to better visualize the distribution of the velocity and flux density of these \meth masers, Figure 
\ref{methfancy} shows the location of the \meth masers as colored circles, where the radii of the circles are 
proportional to the flux density, and the color of the circle indicates the velocity. 
Aside from a few 
bright masers near Sgr C and north of the non-thermal filament known as the Snake, all of the brightest \meth 
masers are also located at positive Galactic longitudes.  

\begin{figure}
	
	\includegraphics[width=0.75\textwidth]{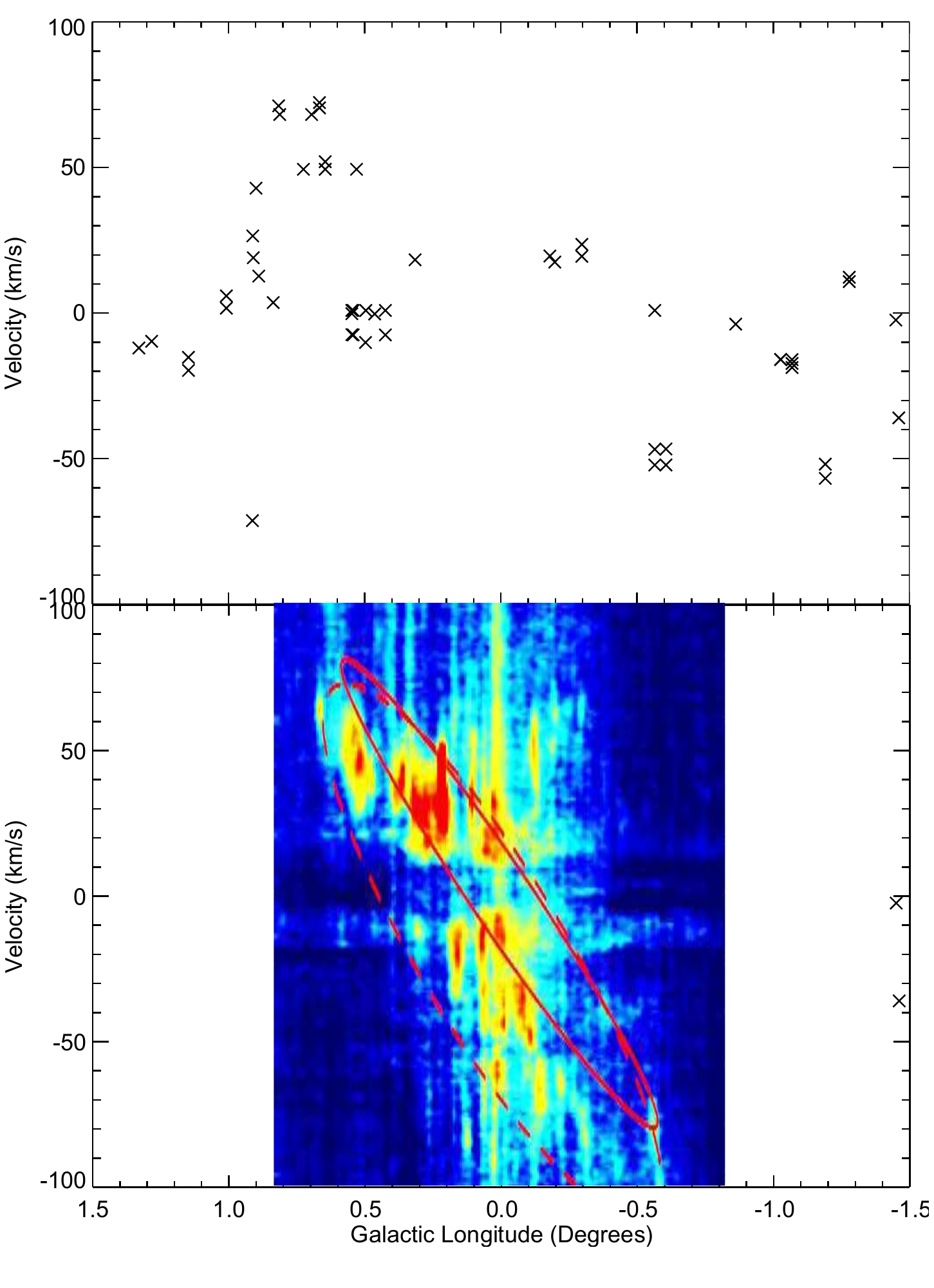}
	\caption{\textbf{Top:} Position - Velocity plot of our detected \meth masers. \textbf{Bottom:} Position - Velocity map of CII from \protect\cite{langer17} for reference (the red ellipses are two different fits to describe the kinematics of the CMZ).}\label{lvmeth}
\end{figure}

In order to better view  the velocity distribution of 
these \meth masers, Figure \ref{lvmeth} shows the velocities of the \meth masers as a function of Galactic longitude (commonly 
referred to as an $l-v$ diagram). Aside from a single maser at G0.912--0.060 (ID 6 in Table \ref{tab:gaus:meth}) with a 
velocity of --71 \kms, all the \meth masers appear to follow the same general trend of the CMZ as shown by a similar $l-v$ diagram of CII from
\cite{langer17}.  

\begin{figure*}
	\includegraphics[width=0.75\textwidth]{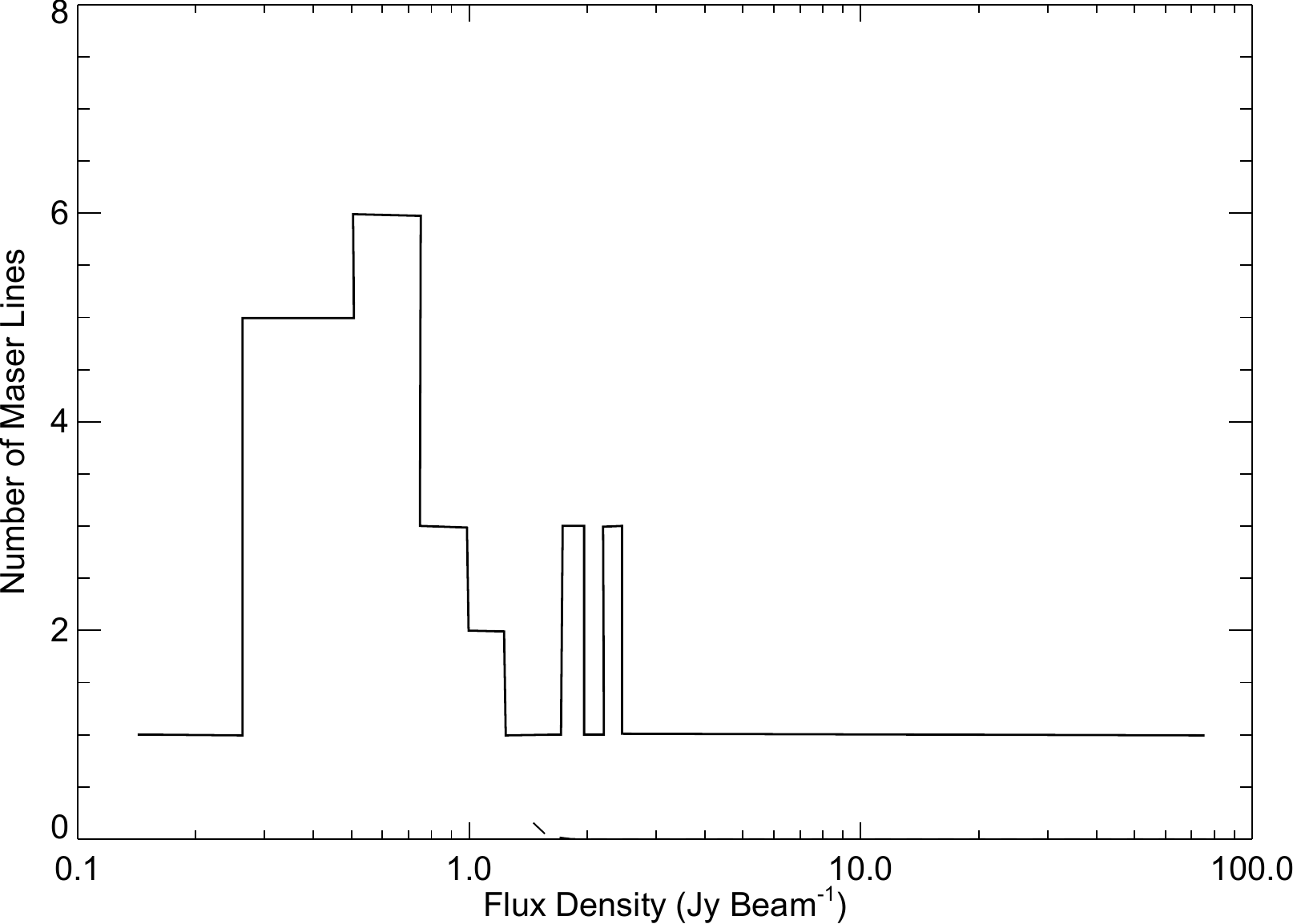}
	\caption{Histogram showing the flux density distribution of our detected \meth masers from Table \ref{tab:gaus:meth}. 
		(bin size of 0.4 Jy beam$^{-1}$, which is $\sim5\sigma$).  }\label{hist:meth}
\end{figure*}

Figure \ref{hist:meth} shows the flux density 
distribution (with a bin size of 0.4 Jy beam$^{-1}$, which is $\sim5\sigma$). Although suffering from a small number 
of detections, fainter ($<$ 1 Jy) masers appear more common. This distribution is similar to a survey done by 
Arecibo (\citealt{olmi14}) finding 
 a majority of 6.7 GHz \meth masers with low flux densities ($<$ 1Jy). 
 These low flux masers have  gone undetected 
by previous surveys. 
It is likely that a population of \meth masers with low flux densities may be located in the CMZ. 

\subsection{New 6.7 GHz \meth Detections}

 We detected a total of 43 \meth masers from 
28 distinct locations. 12 of these locations already had 6.7 GHz \meth masers detected towards them (see footnotes in 
Table \ref{tab:gaus:meth}). This leaves 16 locations with newly detected 6.7 GHz \meth maser candidates that were not 
previously detected by the MMB survey or \cite{caswell09}.

\begin{figure*}

	\includegraphics[width=\textwidth]{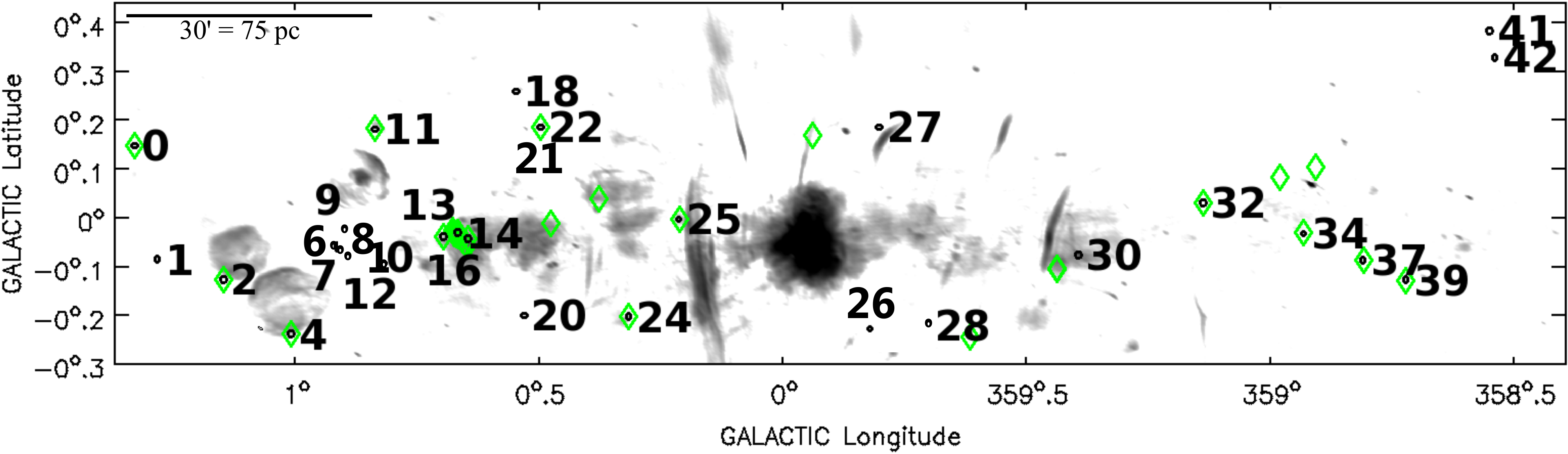}
	\caption{Our detected VLA \meth masers from Table \ref{tab:gaus:meth} (black circles) and the \meth masers from the MMB survey (green diamonds) have been plotted on 90 cm continuum map (\protect\citealt{LawZad08}). We detected masers towards 14 of the 22 locations that the MMB survey detected masers towards.}\label{fig:usvmmb}
\end{figure*}

There are  5 previously reported  \meth masers that are  not detected in our 
survey. Figure \ref{fig:usvmmb} shows the locations of masers we detected 
as well as those detected by the 
MMB survey (\citealt{caswell10}).  The MMB detected masers at Galactic coordinates of 358.906\dee+0.106\dee, 
359.938\dee+0.170\dee, 0.376\dee+0.040\dee, 0.475\dee--0.010\dee, as well as a variety of additional positions near 
Sgr B2 that were not detected in our survey.  
There are prominent imaging artifacts  near the bright \meth masers\footnote{Due to the sources being bright and compact and each field only being observed for short periods of time}, such as Sgr B2, 
and it is therefore likely that we did not include some of the \meth maser 
detections of \cite{caswell10} near Sgr B2 because of confusion with 
these artifacts. We missed the \meth 
maser near 0.378\dee+0.040\dee, which was only faintly detected by the MMB survey (\textless 0.62 Jy, \citealt{caswell10}). We also did not detect the \meth maser \cite{caswell09} detected at 
0.475\dee--0.010\dee.  While it is unclear why this \meth maser was not detected in our survey (their detection had a 
peak of 3.14 Jy near 28.8 \kms), \cite{caswell10} also did not list a \meth maser detected at 0.475\dee--0.010\dee. 

\subsection{Brief Discussions of Specific 6.7 GHz \meth Masers}\label{sec:res:methsum}
The following sections provide brief summaries of some of the 6.7 GHz \meth masers that we detected.  
We give a brief description of the \meth maser spectra followed by comparisons to other previous detections.  We 
compare our detections to those of \cite{caswell09} and the MMB  survey.  We 
then also compared the velocities of these \meth masers to that of HCN (1-0) from \cite{mopra12}.  Finally, we 
compare the locations of these masers to 24 \mum\ (12500 GHz) and 250 \mum\ (1200 GHz) observations (\citealt{far09} 
and \citealt{higal}, respectively).  24 \mum\ emission is produced by HII regions (among other things) and 250 \mum\ emission is produced 
by  dust, which would also indicate the presence of molecular clouds. In the following sections, we 
refer to some of the ATCA detected 22 GHz \wat masers from \cite{thesis}, also see  
\cite{kreiger17a}.  The \wat masers are collisionally 
pumped  and are generally  associated with  YSOs or evolved stars.

{\textbf{G1.329+0.150}}
We identify a single \meth maser  at --12 \kms.   \cite{caswell10} also detected this maser at the same location and velocity. Although this velocity does not quite match with the CMZ velocity for a Galactic longitude of 1.3 \dee(\citealt{langer17}), this \meth maser corresponds with a HCN (1-0) cloud at this same velocity (\citealt{mopra12}).

{\textbf{G1.282-0.084}}
We identify a single \meth maser at --9.7 \kms. This is one of our new detections. It lies near the edge of HCN (1-0) emission at the same velocity and there is 24 \mum\ emission present 
(\citealt{mopra12}; \citealt{far09}). \cite{thesis} and \cite{kreiger17a} also show a new \wat maser candidate towards 
this location with a velocity of --12 \kms.

{\textbf{Sgr D HII Region (G1.147--0.125)}}	
Sgr D is known to be composed of two parts: a northern HII region and a southern super nova remnant (SNR).  While it is unclear if the two sources are physically associated with each other, both have displayed similar molecular line emission, indicating that the may be physically located near each other (\citealt{downes80}). Figure \ref{fig:sgrdmeth} shows a 90 cm continuum map of Sgr D  showing  the northern HII region and the southern SNR components (\citealt{LawZad08}). Over plotted are our detected \meth (+'s) and ATCA \wat (circles) masers (\citealt{kreiger17a}; \citealt{thesis}) along with the corresponding 6.7 GHz \meth spectra. We detect a single \meth maser location just southeast of ($<$ 40\as) the 90 cm emission from  the edge of the northern HII regions with two spectral peaks at --19 and --15 \kms. The MMB survey detected $\sim$ 4 spectral peaks within --14 to --20 \kms. The --15 \kms\ peak coincides with the velocity of one of the ATCA detected \wat masers, supporting that the \wat maser would be associated with high mass star formation (\citealt{kreiger17a}; \citealt{thesis}). Figure \ref{fig:sgrdmeth} also 
shows that this maser is centered towards a clump of 24 \mum\ emission, and all of the mentioned velocities coincide with HCN (1-0) emission towards this location, supporting that this \meth maser is associated with a corresponding molecular cloud (\citealt{mopra12}; \citealt{far09}). 

{\textbf{Sgr D SNR (G1.008--0.237)}}	
G1.008--0.237 is located  south of the 90 cm emission depicting the SNR component of Sgr D 
and is shown in Figure \ref{fig:sgrdmeth} (\citealt{LawZad08}). We detected two spectral peaks at 1.7 and 5.9 \kms. \cite{caswell10} detected similar peaks at $\sim$ 1 and 6 \kms\ although they resolved their 1 \kms\ peak into a double peak. The $\sim$ 1 \kms\ peak matches the 0.7 \kms\ velocity of one of the ATCA \wat masers that was also detected from the same location (\citealt{kreiger17a}; \citealt{thesis}), supporting that it is associated with high mass star formation. There is 250 \mum\ emission towards this location (\citealt{higal}).

{\textbf{G0.912--0.060}}
We identify a new \meth maser  at -71 \kms.  This large negative velocity makes this maser unique, as it can be seen that it stands alone among the \meth masers in the $l-v$ diagram of Figure \ref{lvmeth}.  A comparison to similar $l-v$ diagrams of \cite{langer17} and \cite{molinari11} indicates that this maser is not associate with the CMZ.  Further evidence is the lack of any HCN (1-0) emission at this velocity (\citealt{mopra12}).  This differing velocity is also made evident by how we detected other nearby (within 36 \as) \meth masers (G0.911-0.053 and G0.909-0.061) with much different velocities of 26 and 19 \kms.

{\textbf{G0.9--0.0}}
We identify 4 additional new \meth masers at locations of: G0.911--0.05, G0.909--0.06, G0.899--0.028, G0.889--0.079.  Unlike G0.912--0.060, these all have velocities of: 26, 19, 42, and 12 \kms, respectively, all of which better match the velocities of the CMZ molecular gas. While at the exact locations of these masers, there does not appear to be corresponding HCN (1-0) emission at these velocities, there is corresponding emission nearby (within 50\as, 2 pc).   All the masers located are near (within 3\am) HCN (1-0) emission at their corresponding velocities and 250 \mum\ emission.  G0.911--0.053 is located towards a YSO candidate identified from 4.5 \mum\ excess by \cite{far09}.  The low flux density ($<$0.3 Jy beam$^{-1}$) and narrow channel widths of G0.909--0.061, and G0.889--0.079 could indicate that these are false detections caused by channel noise and could benefit from follow up observations to confirm that they are masers.

{\textbf{G0.836+0.184}}	
We detect a single \meth maser at 3.6 \kms.  \cite{caswell10} previously detected this maser, although they also detected a $\sim$2 \kms\ component that does not show up in our observations. The \meth maser is co-located with 250 \mum\ emission (\citealt{higal}).  This position also matches  the location of a YSO candidate identified by \cite{far09}.

{\textbf{Sgr B2 (G0.695--0.038, G0.666--0.029, and G0.645--0.042)}}	
The Sgr B complex is the 2nd brightest radio source in the CMZ, and is broken into Sgr B1 (G0.506--0.055) and Sgr B2 (G0.667--0.036). Sgr B2 is the most prominent star forming site in the CMZ, with dozens of compact and ultra-compact HII regions (\citealt{mehringer93}). Sgr B2 is broken into three different regions: North (N), Main (M), and South (S), with Sgr B2(M) containing the most HII regions.

We identify \meth masers towards three separate locations (G0.695--0.038, G0.666--0.029, and G0.645--0.042) with velocities spanning 49 to 72 \kms. Figure \ref{fig:sgrb2meth} shows the locations of our \meth maser candidates superimposed on a 90 cm continuum map (\citealt{LawZad08}).  The corresponding spectra are shown as insets. G0.666--0.029 coincides with the location of the molecular cloud Sgr B2(N) and contains two peaks at 70 and 72 \kms\ while G0.645--0.042 coincides with the location of the molecular cloud Sgr B2(S) and contains two peaks at 49 and 52 \kms.  \cite{green15} and the MMB survey (\citealt{caswell10}) identified 6.7 GHz \meth masers from six different locations from G0.665--0.036 to G0.677--0.025.  These locations contain a variety of velocities, but they all are concentrated about 50 and 70 \kms\ (spanning $\sim$ 10 \kms\ about each of these two central velocities), matching with our VLA detections at  G0.666--0.029, and G0.645--0.042. \cite{houghton95} detected a total of 11 maser locations with their ATCA observations that targeted SG B2, and all of their masers had similar velocities spanning 45 to 75 \kms. Many of the additional \meth masers detected by \cite{caswell10} and \cite{houghton95} masers have similar spectra with overlapping velocity features. Such overlapping velocities would have been construed as possible artifacts in our detection method and removed from our final list of masers, which is why our maser catalog does not include as many maser locations.

Our detected \meth maser at G0.695--0.038 has a single velocity at 68 \kms. \cite{green15}, \cite{houghton95}, and \cite{caswell09} also detected a 6.7 GHz \meth maser at this location, with a variety of velocities spanning 10 \kms\ centered at 69 \kms.  This matches with some of the additional features in our spectra for  G0.695--0.038 (Figure \ref{fig:sgrb2meth}) that were not included in the detections, likely either because Gaussians could not be fitted to them or because of the velocity overlap with the features at G0.66--0.029. Also superimposed on Figure \ref{fig:sgrb2meth} are the locations of the ATCA \wat masers (\citealt{kreiger17a}; \citealt{thesis}). It can be seen that G0.666--0.029 coincides with a \wat maser (with a 71 \kms\ velocity, matching that of the \meth maser), while the other two \meth maser are offset from locations of \wat masers. 

{\textbf{G0.530--0.198}}
This is one of our newly detected \meth masers. We detect a single 49 \kms\ maser towards this location. There is 24 \mum\ emission less than 36\as (1.5 pc) away  (\citealt{far09}). The 49 \kms\ velocity matches with the peak velocity of HCN (1-0) emission towards this same location, supporting that this maser is associated with the molecular cloud (\citealt{mopra12}).

{\textbf{G0.49+0.188}}	
We detected \meth masers towards two locations (G0.497+0.188 and G0.496+0.188) with spectra near --7 and 0 \kms. The MMB survey  identified a single location with a variety of spectral peaks spanning --12 to 2 \kms, although their velocities are all located primarily near --9 and --1 \kms.  We are able to resolve these two clumps of spectral lines into two distinct locations located less than 4\as apart, which falls within the resolution of \cite{caswell10}.  G0.49+0.188 coincides with both 24 \mum\ and 250 \mum\ emission (\citealt{far09}; \citealt{higal}). 

{\textbf{G0.315--0.201}}	
We identify a maser at 18 \kms.  The MMB survey detected three 6.7 GHz \meth masers towards this location with velocities near 15, 18 and 20 \kms.  While we only identify the 18 \kms\ maser, there are two low level peaks that occur near 15 \kms\ and 20 \kms\ in our spectra (Figure \ref{fig:G0.315--0.201}). These were not registered as detections though, likely due to the low level. The 18 \kms\ maser velocity matches with the peak velocity of HCN (1-0) towards this location (\citealt{mopra12}). Figure \ref{fig:G0.315--0.201} shows that this maser coincides with 24 \mum\ and 250 \mum\ emission (\citealt{far09}; \citealt{higal}).

{\textbf{G0.212--0.001}}	
We identify a single \meth maser at 49 \kms.  The MMB survey identified  four 4 spectral peaks towards this same location with velocities spanning 41 to 50 \kms.  Additional spectral peaks are detected 
near 42 and 48 \kms, however these peaks are only 1-2 channels wide and can not be fitted by  Gaussians, 
and thus are  not included as detections. There are also multiple nearby (within 15\as) \wat masers (\citealt{kreiger17a}; \citealt{thesis}).  One \wat maser, contains a spectral peak at 48 \kms, matching that of our detected \meth masers, indicating that this \wat maser is associated with high mass star formation. There is HCN (1-0) emission towards this location with velocities spanning 10-95 \kms\ (\citealt{mopra12}). This \meth maser coincides with 24 \mum\ emission (\citealt{far09}).  

{\textbf{G359.703--0.218}}
We identify two \meth masers with velocities of 19 and 23 \kms. There is also a faint peak near 25 \kms\ evident in our spectra.  However this feature did not make it into our detections due to its narrow channel width.  The MMB survey identified multiple \meth masers towards this location with velocities spanning 14 to 27 \kms, although the MMB survey identified this \meth maser at a slightly different location (359.615\dee -0.243\dee). It is unclear the reason for this $\sim$6\am difference in location.  
There is corresponding HCN (1-0) emission at these 14-27 \kms\ velocities indicating that these masers are likely associated with the molecular gas (\citealt{mopra12}).  This maser is located within 1.2\am (3 pc) of 24 \mum\ emission (\citealt{far09}). 

{\textbf{Sgr C (G359.394--0.075)}}	
Sgr C is known to contain a shell-like HII region located within a cavity of molecular gas (\citealt{LisztSpiker1995}).  We identify two \meth masers towards a single location of the Sgr C HII region with velocities at --52 and --46 \kms.  The MMB survey separated these two \meth masers as coming from two locations separated by 7\as. These velocities match with those of 22 GHz \wat masers that have been detected towards this location (\citealt{caswell11}; \citealt{taylor93}), and are near the -60 \kms\  velocity of the nearby molecular gas (\citealt{LisztSpiker1995}). 

{\textbf{G358.931--0.030}}	
We identify three \meth masers towards a single location with velocities at --18, --17, and --16 \kms.  There are also fainter ($<$ 0.1 Jy beam$^{-1}$) peaks (near --20, --19, and --15 \kms) that did not qualify as detections due to the low levels and narrow channel widths (the --15 \kms\ peak is a single channel wide). The MMB survey identified many \meth masers from this single location with velocities spanning --22 to --14 \kms. There is faint ($<$ 600 MJy/sr) 250 \mum\ emission coincident with this maser location (\citealt{higal}).

{\textbf{G358.809--0.085}}	
We identify \meth masers with velocities at --51 \kms\ and --56 \kms towards a single location, although the peak at --56 \kms\ displays a double peak spectra. The MMB survey also identified multiple \meth masers towards this location with velocities spanning --60 to --50 \kms, which our observations agree with except for their lower level ($<$ 0.5 Jy) detection near --60 \kms, which does not show up in our observations. There is also faint ($<$ 800 MJy / sr) 250 \mum\ emission coincident with these masers (\citealt{higal}).

{\textbf{G358.721--0.126}}	
We identify \meth masers with velocities at 10 and 12 \kms towards a single location. The MMB survey also identified these \meth masers with matching velocities of 10 and 12 \kms. \cite{caswell10} indicated that this velocity places the masers at a distance $<$ 13.5 kpc and outside the solar circle, potentially placing these masers in the CMZ. There is also faint ($<$ 800 MJy / sr) 250 \mum\ emission coincident with these masers (\citealt{higal}).

\section{Conclusions}

Using the VLA, we conducted a blind survey of 6.7 GHz \meth masers covering the inner $\sim3$\dee$\times\ 0.5$\dee\ of the GC. This maser is an exclusive indicator of high mass star formation.  
We detect 43 \meth masers towards 28 distinct locations. $\sim$16 of these are new detections, including a new group of masers located towards G0.9--0.0. We also detect \meth masers towards Sgr B, Sgr C, Sgr D, and other molecular clouds that agree with previous 6.7 GHz \meth masers detected towards these regions. A majority of the masers have velocities consistent with being in the CMZ. The spatial distribution of \meth masers appears to show 
an  asymmetry, with 24 \meth masers detected at
 positive Galactic longitudes and only 10 towards negative longitudes. 
This is similar to how about two-thirds of the molecular gas in this region is located at positive Galactic longitudes. This work demonstrates the benefit of using blind interferometric surveys of masers to identify new sites of star formation towards the CMZ.

\section*{Acknowledgements}
This work was partially supported by the grant AST-0807400 from the NSF, the 
National Radio Astronomy 
Observatory's summer student program and Grote Reber fellowship. 

\vspace{5mm}

\bibliography{./Ref}

\clearpage

\begin{table}
	
	\caption{VLA \meth Survey  Observation}\label{tab:obs:meth}
	\begin{tabular}{l | c l}
		\hline
		
		VLA Configuration & & A \\
		Observing Date & & 2015 July 09 \\
		Angular Resolution & & 0.86\as $\times$ 0.29\as (P.A. $13^\circ$) \\
		Primary Beam FWHM & & 6.75\am \\
		Spectral Resolution & &   0.4 km s$^{-1}$ (8.06 kHz) \\
		Velocity Coverage & &  \+ 390 km s$^{-1}$ \\
		Average RMS & & 0.08 Jy beam$^{-1}$ per channel \\
		Total Observing Time & & 3 hours \\
	\end{tabular}
\end{table}


\begin{table}
	
	\caption{Comparison of untargeted CMZ 6.7 GHz \meth Maser Surveys}\label{tab:methsurv:comp}
	\begin{tabular}{l | c c c }
		\hline
		& Rickert2018 & MMB $^{a}$ & Caswell96\\
		\hline
		\hline
		Resolution$^{b}$ & 0.86\as & 0.4\as/24\as & 2\as $\times$ 4\as \\
		Channel Width  (\kms)& 0.4  & 0.11/0.4  & 0.18 \\
		Sensitivity & 0.08 Jy beam$^{-1}$ & 0.07 Jy/0.17 Jy & 0.16 Jy  \\
\hline

\footnote{$^{a}$The 2$^\mathrm{nd}$ value refers to the initial single dish observations, while the 1$^\mathrm{st}$ value is based off of the the later ATCA follow up observations.
$^{b}$These resolutions are not completely comparable, as they refer to the beam fwhm, source positional accuracy, and beam full width half power (fwhp), respectively.
}
\end{tabular}
\end{table}

\begin{landscape}

	\begin{table*}
				\caption{\meth Maser Spectra Gaussian Fits}\label{tab:gaus:meth}
\resizebox{\textwidth}{!}{			
			\begin{tabular}{c c c c r c r r r r l}

		\hline
ID & Source & RA. & Dec. & Peak Flux & RMS & V$_{LSR}$ & FWHM & Integrated Flux & Peak Brightness & \multicolumn{1}{c}{Associated} \\
 & ($l,b\  ^\circ$) & $^{h}\ ^{m}\ ^{s}$  & \dee\ \am \as  &  Density$^{k}$ &  &  (km s$^{-1}$) &  (km s$^{-1}$) & \multicolumn{1}{c}{(Jy s km$^{-1}$} & Temperature & \multicolumn{1}{c}{Sources} \\
 &  & \multicolumn{2}{c}{(J2000.0)}& \multicolumn{2}{c}{(Jy beam$^{-1}$)} & & &  \multicolumn{1}{c}{beam$^{-1}$)} & ($\times10^3$ K) & \\
		\hline
		\hline
0  &  1.329 $+$0.150   &  17 48 10.301   &  $-$27 43 20.877   &  0.72 \+ 0.14   &  0.04  &   $-$12.0 \+$<$ 0.1   &   0.4 \+ 0.1   &   0.34 \+ 0.10 &  32.4 \+ 1.8  &   $^{b}$ \\
1  &  1.282 $-$0.084   &  17 48 58.080   &  $-$27 52 58.600   &  0.36 \+ 0.02   &  0.01  &   $-$9.7 \+$<$ 0.1   &   0.6 \+$<$ 0.1   &   0.22 \+ 0.02  &  16.2 \+ 0.5   &  $^{j}$ $^{*}$ \\
2  &  1.147 $-$0.125   &  17 48 48.549   &  $-$28 01 11.102   &  0.24 \+ 0.06   &  0.03  &   $-$19.7 \+ 0.3   &   2.6 \+ 0.8   &   0.50 \+ 0.19 &  10.8 \+ 1.4  &  Sgr D HII $^{a}$ $^{b}$  $^{j}$ \\
3  &  ''   &  ''   &  ''   &  15.0 \+ 1.3   &  0.09  &   $-$15.1 \+$<$ 0.1   &   0.4 \+ 0.1   &   6.03 \+ 0.67 &  675.4 \+ 4.1  &  $^{j}$ \\ 
4  &  1.008 $-$0.237   &  17 48 55.287   &  $-$28 11 48.192   &  8.05 \+ 0.19   &  0.14  &   1.7 \+$<$ 0.1   &   0.5 \+$<$ 0.1   &   4.05 \+ 0.15  &  362.4 \+ 6.3 & Sgr D SNR $^{a}$ $^{b}$ $^{j}$ \\
5  &  ''   &  ''   &  ''   &  1.10 \+ 0.06   &  0.06  &   5.9 \+$<$ 0.1   &   0.7 \+$<$ 0.1   &   0.84 \+ 0.07  &  49.5 \+ 2.7 &  $^{j}$ \\
6  &  0.912 $-$0.060   &  17 48 00.435   &  $-$28 11 16.945   &  0.11 \+ 0.02   &  0.02  &   $-$71.3 \+ 0.1   &   0.6 \+ 0.1   &   0.07 \+ 0.02 &  5.0 \+ 0.9  &   $^{*}$\\
7  &  0.911 $-$0.053   &  17 47 58.728   &  $-$28 11 04.789   &  0.12 \+ 0.02   &  0.02  &   26.5 \+ 0.1   &   0.4 \+ 0.1   &   0.05 \+ 0.02 &  5.4 \+ 0.9  &   $^{c}$ $^{*}$ \\
8  &  0.909 $-$0.061   &  17 48 00.205   &  $-$28 11 27.186   &  0.11 \+ 0.02   &  0.02  &   19.0 \+ 0.1   &   0.4 \+ 0.1   &   0.05 \+ 0.02 &  5.0 \+ 0.9  &  $^{*}$ \\
9  &  0.899 $-$0.028   &  17 47 51.201   &  $-$28 10 54.390   &  0.12 \+ 0.04   &  0.02  &   42.9 \+ 0.1   &   0.4 \+ 0.2   &   0.05 \+ 0.02 &  5.4 \+ 0.9  & $^{*}$  \\
10  &  0.889 $-$0.079   &  17 48 01.828   &  $-$28 13 02.861   &  0.12 \+ 0.02   &  0.02  &   12.7 \+ 0.1   &   0.5 \+ 0.1   &   0.03 \+ 0.03  &  5.4 \+ 0.9 & $^{*}$  \\
11  &  0.836 $+$0.184   &  17 46 52.826   &  $-$28 07 35.439   &  3.14 \+ 0.05   &  0.07  &   3.6 \+$<$ 0.1   &   0.4 \+$<$ 0.1   &   1.24 \+ 0.04 &  141.4 \+ 3.2  &   $^{b}$ $^{c}$ $^{d}$ \\
12  &  0.816 $-$0.091   &  17 47 54.383   &  $-$28 17 09.354   &  0.08 \+ 0.01   &  0.01  &   71.2 \+$<$ 0.1   &   0.6 \+ 0.1   &   0.05 \+ 0.02  &  3.6 \+ 0.5  & $^{*}$  \\ 
13  &  0.695 $-$0.038   &  17 47 24.741   &  $-$28 21 43.297   &  11.84 \+ 0.85   &  0.75  &   68.2 \+ 0.1   &   1.7 \+ 0.1   &   18.90 \+ 3.70 &  533.1 \+ 33.8  &   Sgr B2 $^{b}$ $^{c}$ $^{d}$ \\
14  &  0.666 $-$0.029   &  17 47 18.655   &  $-$28 22 54.493   &  24.50 \+ 6.40   &  1.70  &   70.5 \+$<$ 0.1   &   0.4 \+ 0.1   &   13.1 \+ 5.3 &  1103.1 \+ 76.5  &  Sgr B2 $^{a}$ $^{b}$ $^{c}$ $^{d}$ $^{j}$ \\
15  &  ''   &  ''   &  ''   &  5.20 \+ 1.30   &  0.39  &   72.3 \+$<$ 0.1   &   0.6 \+ 0.2   &   3.5 \+ 1.4  &  234.1 \+ 17.6 &  $^{j}$ \\ 
16  &  0.645 $-$0.042   &  17 47 18.654   &  $-$28 24 24.733   &  36.10 \+ 2.30   &  1.80  &   49.5 \+$<$ 0.1   &   1.2 \+ 0.1   &   47.8 \+ 4.6  &  1625.4 \+ 81.0 &   Sgr B2 $^{b}$ $^{c}$ $^{d}$ \\
17  &  ''   &  ''   &  ''   &  6.82 \+ 0.97   &  0.70  &   52.0 \+$<$ 0.1   &   0.8 \+ 0.1   &   11.6 \+ 9.8  &  307.1 \+ 31.5 &   \\ 
18  &  0.547 $+$0.261   &  17 45 54.060   &  $-$28 20 00.412   &  2.36 \+ 0.15   &  0.03  &   $-$0.2 \+ 0.1   &   2.2 \+ 0.2   &   5.47 \+ 0.68 &  106.3 \+ 1.4  & $^{*}$  \\
19  &  ''   &  ''   &  ''   &  0.56 \+ 0.12   &  0.07  &   $-$7.5 \+ 0.1   &   0.7 \+ 0.2   &   0.40 \+ 0.19 &  25.2 \+ 3.2  &   \\ 
20  &  0.530 $-$0.198   &  17 47 38.819   &  $-$28 35 07.443   &  0.21 \+ 0.02   &  0.02  &   49.4 \+ 0.1   &   1.6 \+ 0.1   &   0.34 \+ 0.04 &  9.5 \+ 0.9  &  $^{*}$ \\
21  &  0.497 $+$0.188   &  17 46 03.989   &  $-$28 24 49.675   &  4.03 \+ 0.64   &  0.16  &   $-$10.1 \+ 0.4   &   0.6 \+ 0.1   &   2.48 \+ 0.64  &  181.4 \+ 7.2 &   $^{b}$ $^{c}$ \\
22  &  0.496 $+$0.188   &  17 46 04.000   &  $-$28 24 51.899   &  76.10 \+ 6.60   &  0.70  &   0.9 \+$<$ 0.1   &   0.5 \+ 0.1   &   39.60 \+ 5.40  &  3426.4 \+ 31.5 &   \\
23  &  ''   &  ''   &  ''   &  2.77 \+ 0.41   &  0.09  &   $-$7.4 \+ 0.1   &   1.2 \+ 0.2   &   3.46 \+ 0.80  &  124.7 \+ 4.1 &   \\ 
24  &  0.315 $-$0.201   &  17 47 09.114   &  $-$28 46 16.009   &  14.60 \+ 1.10   &  0.59  &   18.3 \+$<$ 0.1   &   0.6 \+ 0.1   &   8.7 \+ 1.1 &  657.4 \+ 26.6  &   $^{a}$ $^{b}$ $^{c}$ $^{d}$ \\
25  &  0.212 $-$0.001   &  17 46 07.678   &  $-$28 45 20.390   &  1.056 \+ 0.08   &  0.06  &   49.3 \+$<$ 0.1   &   1.05 \+ 0.1   &   1.19 \+ 0.14 &  7.7 \+ 0.5  &  $^{b}$ $^{j}$ \\
26  &  359.820 $-$0.225   &  17 46 04.183   &  $-$29 12 25.540   &  0.17 \+ 0.02   &  0.01  &   19.6 \+$<$ 0.1   &   0.5 \+ 0.1   &   0.09 \+ 0.021 &  10.4 \+ 1.4  &  $^{*}$ \\
27  &  359.802 $+$0.188   &  17 44 24.806   &  $-$29 00 22.806   &  0.23 \+ 0.03   &  0.03  &   17.5 \+ 0.1   &   0.5 \+ 0.1   &   0.11 \+ 0.02 &  339.0 \+ 10.8  & $^{*}$  \\
28  &  359.703 $-$0.218   &  17 45 45.811   &  $-$29 18 10.446   &  7.53 \+ 0.40   &  0.24  &   19.5 \+$<$ 0.1   &   0.5 \+$<$ 0.1   &   3.87 \+ 0.33 &  30.2 \+ 3.2  &   \\
29  &  ''   &  ''   &  ''   &  0.67 \+ 0.19   &  0.07  &   23.6 \+ 0.4   &   2.9 \+ 1.0   &   2.08 \+ 0.94  &  101.8 \+ 7.2  &   \\ 
30  &  359.394 $-$0.075   &  17 44 27.909   &  $-$29 29 31.299   &  0.68 \+ 0.04   &  0.09  &   $-$52.2 \+$<$ 0.1   &   1.0 \+$<$ 0.1   &   0.70 \+ 0.06 &  503.4 \+ 30.6  &   Sgr C \\
31  &  ''   &  ''   &  ''   &  11.18 \+ 0.05   &  0.68  &   $-$46.7 \+$<$ 0.1   &   0.5 \+ 0.10   &   5.78 \+ 0.04 &  594.3 \+ 19.8  &   \\ 
32  &  359.138 $+$0.032   &  17 43 25.672   &  $-$29 39 17.144   &  13.20 \+ 1.00   &  0.44  &   $-$3.8 \+$<$ 0.1   &   0.8 \+$<$ 0.1   &   10.62 \+ 0.60 &  41.4 \+ 2.7  &   $^{a}$ $^{b}$ $^{c}$ $^{h}$ \\
33  &  358.973 $+$0.088   &  17 42 48.433   &  $-$29 45 55.095   &  0.27 \+ 0.04   &  0.05  &   $-$16.0 \+$<$ 0.1   &   0.6 \+ 0.1   &   0.18 \+ 0.04   &  78.3 \+ 2.3 &   \\ 
34  &  358.931 $-$0.030   &  17 43 10.031   &  $-$29 51 45.663   &  1.74 \+ 0.22   &  0.05  &   $-$18.7 \+$<$ 0.1   &   0.5 \+ 0.1   &   1.2 \+ 1.9 &  53.1 \+ 1.8  &   $^{b}$ \\
35  &  ''   &  ''   &  ''   &  1.18 \+ 0.14   &  0.04  &   $-$17.3 \+$<$ 0.1   &   0.7 \+ 0.1   &   1.1 \+ 1.4  &  139.6 \+ 4.1 &   \\ 
36  &  ''   &  ''   &  ''   &  3.10 \+ 0.15   &  0.09  &   $-$16.0 \+$<$ 0.1   &   0.5 \+$<$ 0.1   &   1.52 \+ 0.50  &  83.3 \+ 2.7 &   \\ 
37  &  358.809 $-$0.085   &  17 43 05.408   &  $-$29 59 45.583   &  1.85 \+ 0.21   &  0.06  &   $-$56.8 \+ 0.1   &   1.9 \+ 0.2   &   3.78 \+ 0.25 &  9.0 \+ 0.9  &   $^{b}$ \\
38  &  ''   &  ''   &  ''   &  0.20 \+ 0.32   &  0.02  &   $-$51.9 \+ 0.6   &   0.8 \+ 1.5   &   0.18 \+ 0.43  &  27.0 \+ 1.8 &   \\ 
39  &  358.721 $-$0.126   &  17 43 02.312   &  $-$30 05 29.942   &  0.60 \+ 0.02   &  0.04  &   10.8 \+$<$ 0.1   &   1.0 \+$<$ 0.1   &   0.62 \+ 0.04 &  15.3 \+ 0.9  &   $^{b}$ \\
40  &  ''   &  ''   &  ''   &  0.34 \+ 0.02   &  0.02  &   12.3 \+$<$ 0.1   &   1.3 \+ 0.1   &   0.47 \+ 0.06 &  12.6 \+ 1.4  &   \\ 
41  &  358.550 $+$0.384   &  17 40 37.160   &  $-$29 58 03.303   &  0.28 \+ 0.04   &  0.03  &   $-$2.4 \+ 0.1   &   0.4 \+ 0.1   &   0.12 \+ 0.03  &  12.2 \+ 1.4 & $^{*}$  \\
42  &  358.539 $+$0.330   &  17 40 48.193   &  $-$30 00 20.446   &  0.27 \+ 0.06   &  0.03  &   $-$36.0 \+ 0.4   &   0.1 \+$<$ 0.1   &   0.01 \+ 0.03 &  1.8 \+ 1.1  &  $^{*}$ \\
\hline
\multicolumn{5}{l}{$^{a}$\wat masers: \citealt{walsh14}; 30\as} & \multicolumn{5}{l}{$^{h}$SiO masers: \citealt{harju98}; 57\as}\\
	\multicolumn{5}{l}{$^{b}$6.7 GHz \meth masers: \citealt{caswell10}; 20\as} & \multicolumn{5}{l}{$^{i}$36 GHz \meth masers: \citealt{cotton16}; 0.86\as}\\
	\multicolumn{5}{l}{$^{c}$4.5\mum\ YSO candidates: \citealt{far09}; 6\as} & \multicolumn{5}{l}{$^{j}$\wat masers: \citealt{kreiger17a}; 26\as}\\
	\multicolumn{5}{l}{$^{d}$6.7 GHz \meth masers \citealt{caswell09}; 2\as} & \multicolumn{5}{l}{$^{k}$The Flux density, V$_{LSR}$, and FWHM are all from Gaussian fits,}\\
	\multicolumn{5}{l}{$^{e}$22 GHz \wat and 1612 OH masers \citealt{lorant02}; 2\as} &\multicolumn{5}{l}{ with fitting uncertainties.} \\
	\multicolumn{5}{l}{$^{f}$22 GHz \wat and 1665 MHz OH masers \citealt{forster99}; 2\as} & \multicolumn{5}{l}{$^{*}$ New detections} \\
	\multicolumn{5}{l}{$^{g}$OH/IR stars: \citealt{lorant98}; 2.5\as} & \\
\end{tabular}
}
\end{table*}
\end{landscape}

\begin{figure*}
	\includegraphics[width=\textwidth]{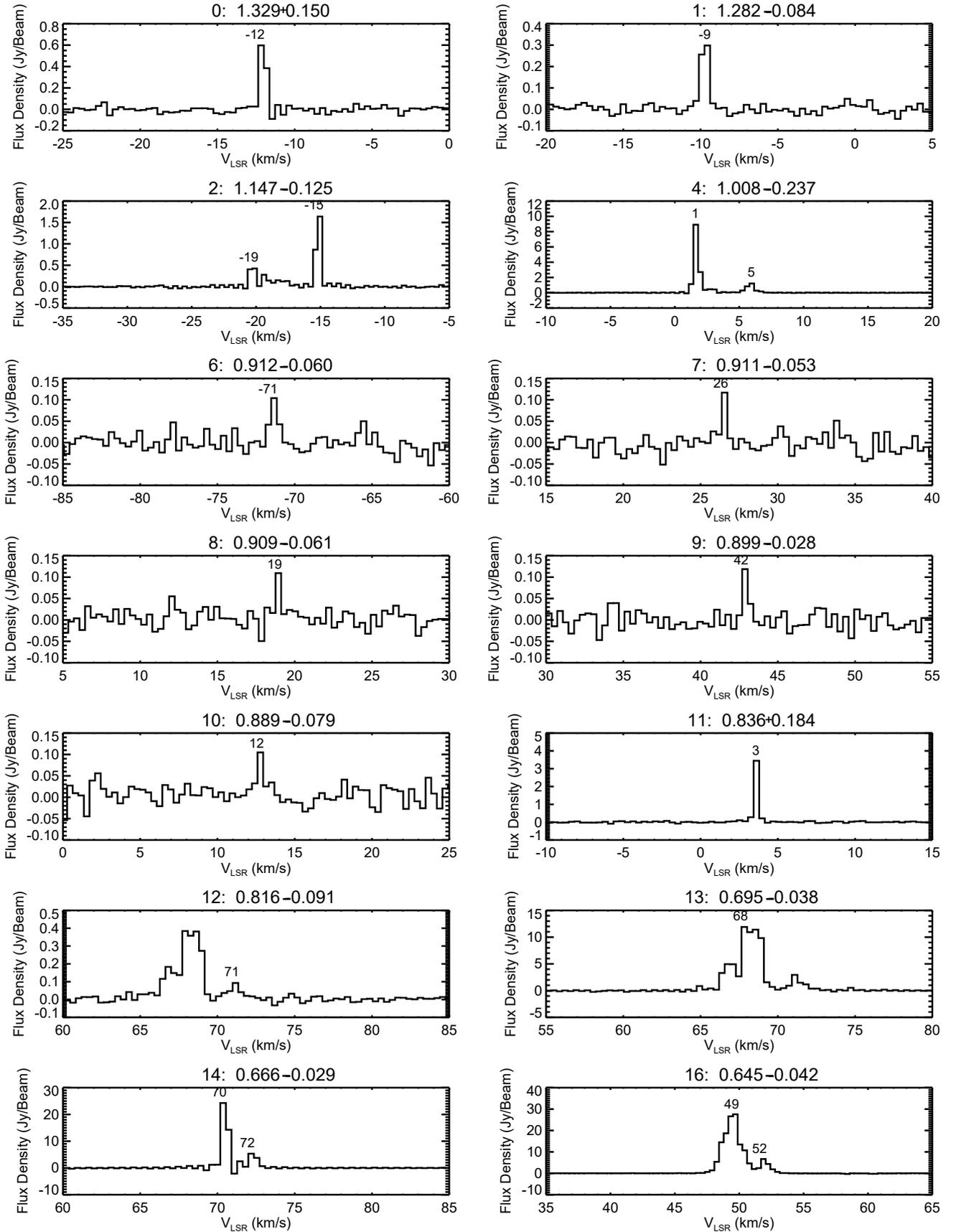}
	\caption{27 spectra for all the 42 potential maser candidates from the blind  VLA 6.7 GHz \meth survey. The header gives the maser number (from Table \ref{tab:gaus:meth} some masers are co-located; in these cases only the first maser number is listed) and the position in Galactic coordinates.  The channel width is 0.4 \kms.}\label{methspec}
\end{figure*}
\begin{figure*}
	\includegraphics[width=\textwidth]{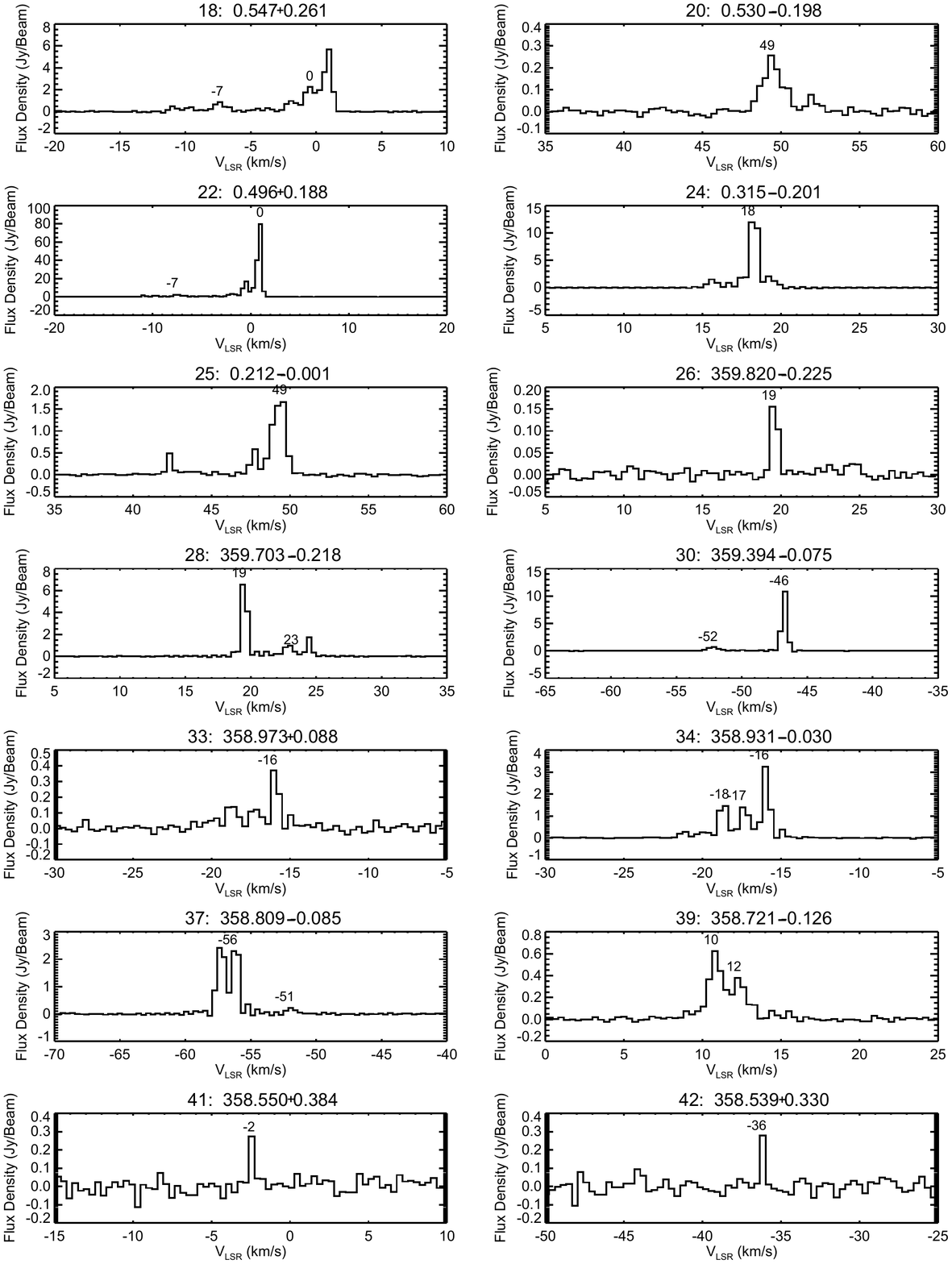}
		\contcaption{Figure \ref{methspec}}
\end{figure*}


\begin{figure*}
	\includegraphics[width=0.75\textwidth]{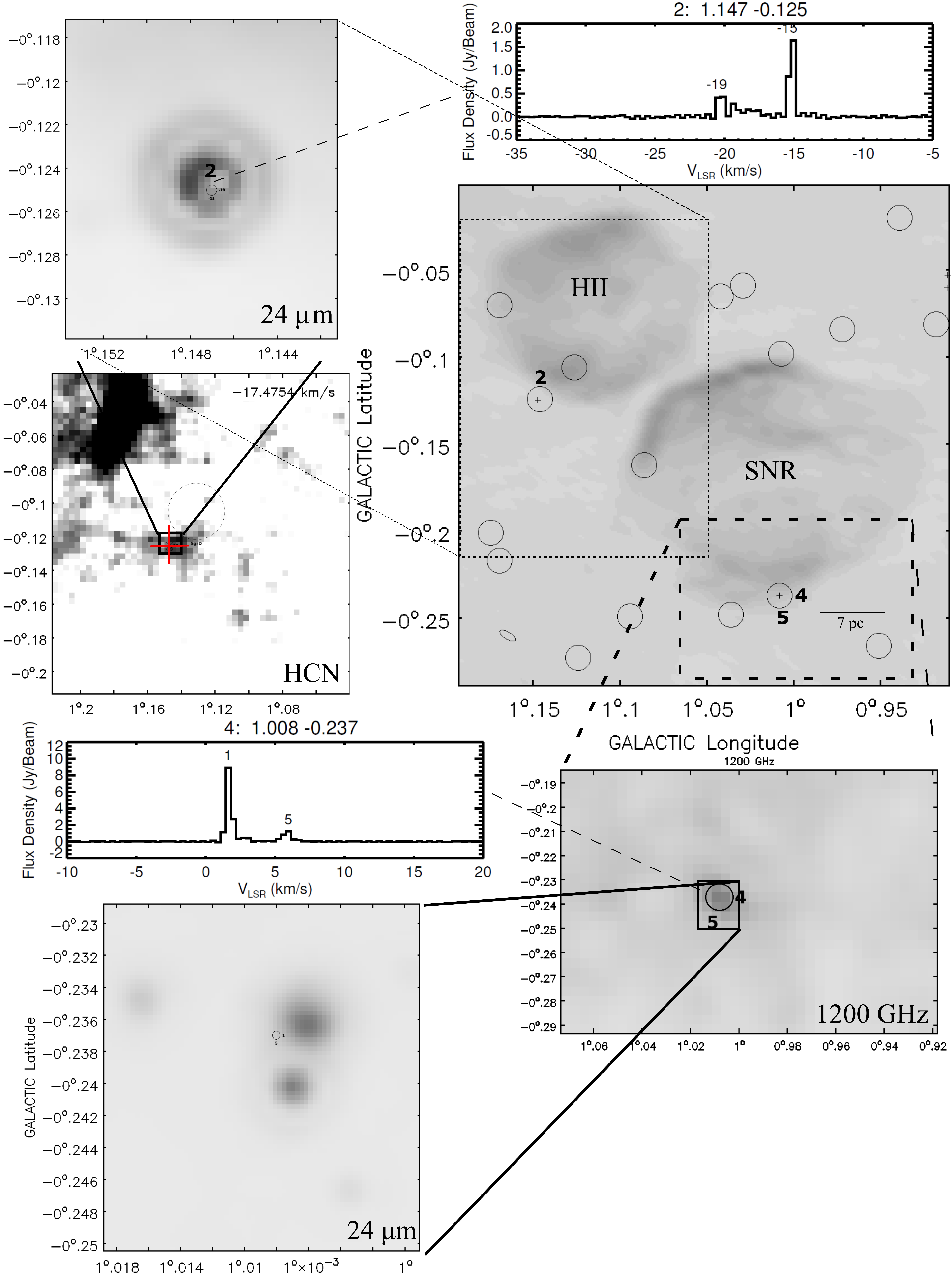}
	\caption{A 90 cm map of Sgr D (\protect\citealt{LawZad08}) with the detected ATCA \wat (circles,\protect\citealt{kreiger17a}; \protect\citealt{thesis}) and our VLA \meth masers plotted (+'s). The northern HII region and the southern SNR are quite evident in the 90 cm map. Surrounding are the corresponding \meth spectra from Figure \ref{methspec}. The insets correspond to: an HCN channel map (matching the corresponding \meth maser velocity, 24 \mum\ and 1200 GHz (250 \mum) maps (\protect\citealt{mopra13}; \protect\citealt{far09}; \protect\citealt{higal}).}\label{fig:sgrdmeth}
\end{figure*}

\begin{figure*}
	
	\includegraphics[width=\textwidth]{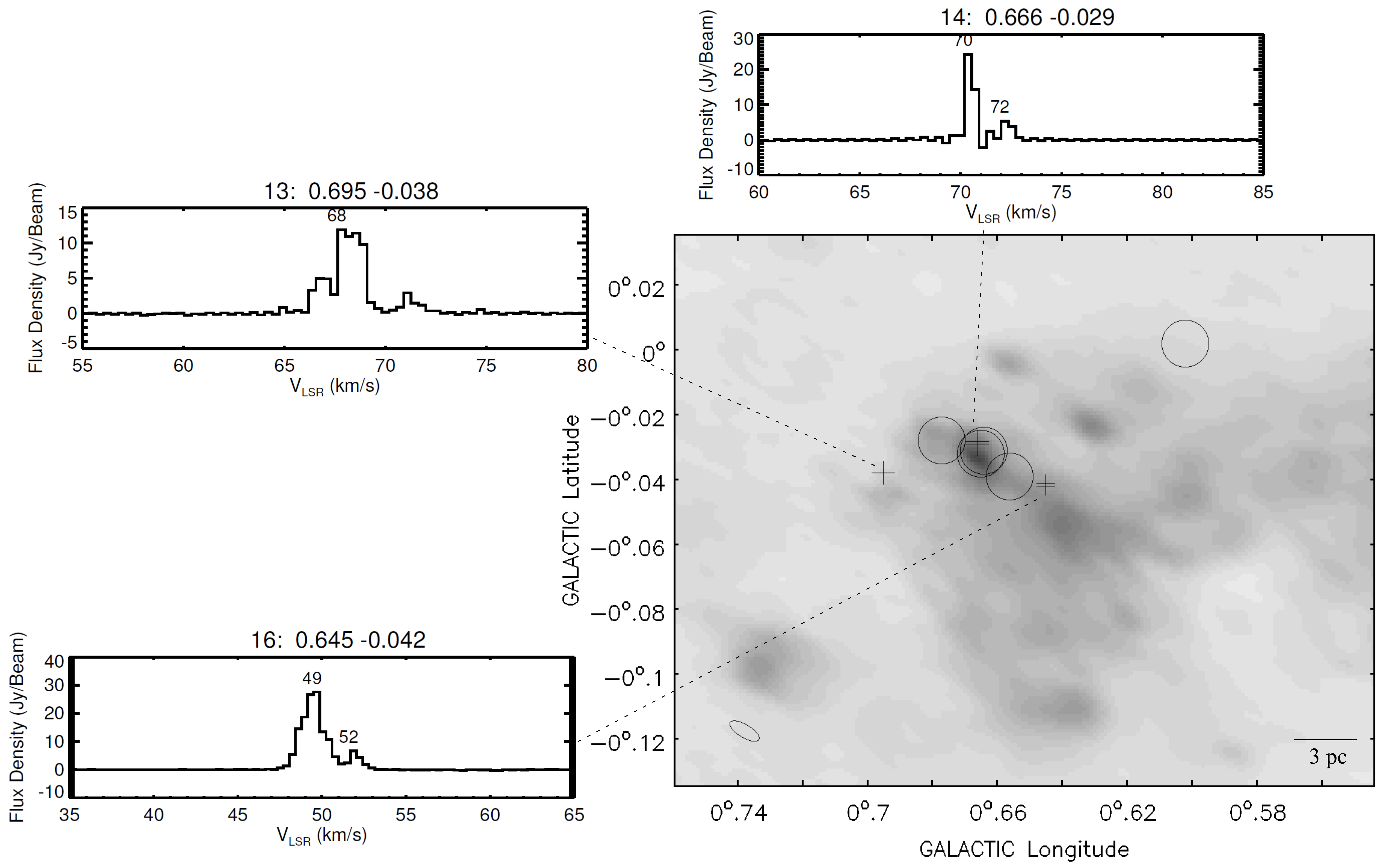}
	\caption{A 90 cm map of Sgr B2 (\protect\citealt{LawZad08}) with detected ATCA \wat (circles, \protect\citealt{kreiger17a}; \protect\citealt{thesis}) and our \meth masers plotted (+'s). Surrounding are the corresponding \meth spectra from Figure \ref{methspec}.}\label{fig:sgrb2meth}
\end{figure*}


\begin{figure*}
	\includegraphics[width=\textwidth]{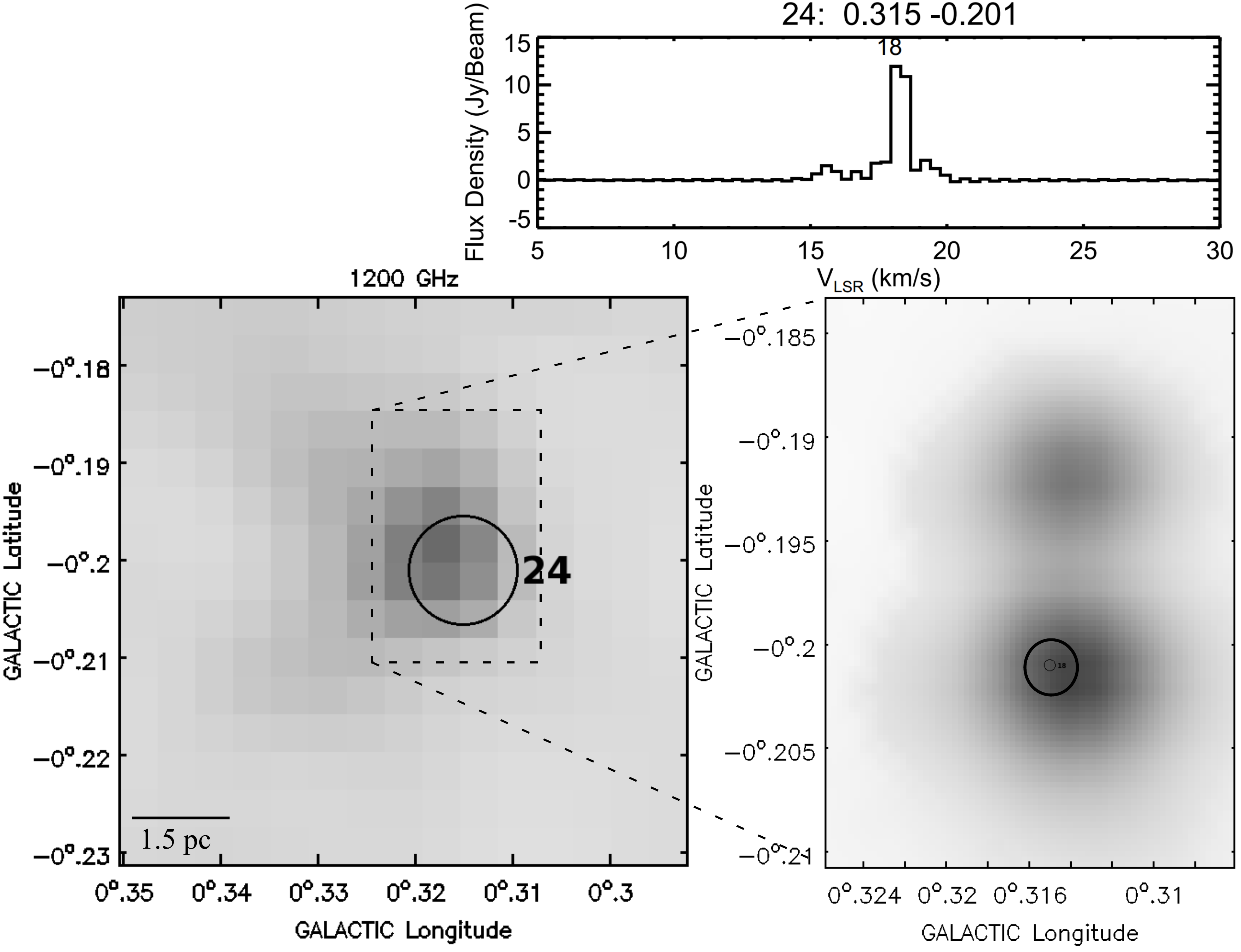}
	\caption{250 \mum\ and 24 \mum\ (inset) map for G0.315--0.201 (with the corresponding \meth spectrum from Figure \ref{methspec} shown near the top).  The position of the 18 \kms\ 6.7 GHz \meth maser is labeled with a circle (\protect\citealt{higal}; \protect\citealt{far09}).}\label{fig:G0.315--0.201}
\end{figure*}

\bsp	
\label{lastpage}
\end{document}